\documentclass[aps,prf,showpacs]{revtex4-1}

\usepackage{amsmath,amssymb,graphicx,graphics,bm}
\usepackage{hyperref}
\usepackage{graphics,graphicx}
\usepackage{amsmath,amssymb}

\begin{document}

\title{Generalized fluctuation-dissipation relations in confined geometries
and concentrated conditions}
\author{Massimiliano Giona$^*$, Giuseppe Procopio and  Chiara Pezzotti}

\affiliation{Dipartimento di Ingegneria Chimica, Materiali, Ambiente La Sapienza Universit\`a di Roma\\ Via Eudossiana 18, 00184 Roma, Italy\\
E-mail: massimiliano.giona@uniroma1.it  \\
$^*$corresponding author}

\date{\today}

\begin{abstract}
This article extends the fluctuation-dissipation analysis to generic
complex fluids in confined geometries and to  all the cases the
hydromechanic fluid-interaction kernels may depend on  the particle
position.  
This represents a completely new way of enforcing fluctuation-dissipation
theory just because the primary target is to derive an explicit
functional expression for the hydromechanic  force (that is
unavailable from linear hydrodynamic theory) from fundamental
thermodynamic principles at equilibrium (while in the
classical Kubo theory the  memory kernels are explicitly
known, stemming from the mean-field hydromechanics of  fluid-particle
interactions). In this way, either the representation of hydromechanic interactions and the explicit representation of the thermal forces are derived at the
same time from thermodynamic principles.
The physical and conceptual implications of these results are     
addressed.
The theory can be extended to concentrated conditions and to suspensions,
as well as to active particle in confined geometries accounting for the
most general linear fluid-dynamic conditions.
\end{abstract}

\maketitle

\noindent
\section{Introduction}  The fluctuation-dissipation theory represents
a milestone in statistical physics with relevant implications
in all the branches of physics.
The theory originates from the Einstein's analysis of Brownian
motion in a fluid at constant temperature $T$, and 
connects the phenomenological parameter associated  with the
intensity of fluctuations (the diffusion coefficient $D$) to the
strength of dissipation  expressed by the friction factor $\eta$,
(accounting from particle's  hydromechanics
in a Newtonian fluid under instantaneous Stokes conditions), thus ultimately
obtaining the   well-known Stokes-Einstein relation $D \, \eta=k_B \, T$,
where $k_B$ is the Boltzmann constant \cite{gen1,gen2}.

The equations of motion of a particle of mass $m$ and velocity ${\bf v}$
 in a stagnant fluid
at constant temperature 
can be expressed as
\begin{equation}
m \, \frac{d {\bf v}(t)}{d t}= {\bf F}_{f \rightarrow p}[{\bf v}(t)] +  {\bf R}(t)
\label{eq1}     
\end{equation}
where ${\bf F}_{f \rightarrow p}[{\bf v}(t)]$ is the mean-field contribution to
the force exerted by the fluid onto the particle deriving from linear hydrodynamics, and ${\bf R}(t)$
is the thermal fluctuation force. ${\bf F}_{f \rightarrow p}[{\bf v}(t)]$ 
is a linear stationary, causal, and dissipatively stable 
functional of the particle velocity.

A major generalization of the theory came with the work by
Kubo \cite{kubo1,kubo2}  considering more general hydromechanic interactions possessing
a memory character, and thus addressing the equilibrium property
of  fluid-particle interactions driven by Generalized
Langevin Equations (GLE), in which 
the force ${\bf F}_{f \rightarrow p}[{\bf v}(t)]$
attains
the form
\begin{equation}
{\bf F}_{f \rightarrow p}[{\bf v}(t)] = \int_0^t {\bf h}(t-\tau) \, {\bf v}(\tau) \, d \tau
\label{eq2}
\end{equation}
where ${\bf h}(t)$ is a tensorial dissipative memory kernel.

  Following Kubo \cite{kubo1,kubo2}, a distinction
has been introduced between the fluctuation-dissipation relations (theorems)
of the first  and second kind, (henceforth referred to as FD1k, FD2k),
connecting  the autocorrelation properties
of the particle velocity  ${\bf v}(t)$ and of the fluctuation force ${\bf R}(t)$ to the structure of the memory kernel ${\bf h}(t)$, respectively.
The Kubo theory is essentially grounded on two basic assumptions:
i) the Langevin condition  \cite{langevin}
\begin{equation}
\langle R_i(t) \,v_j(0) \rangle_{\rm eq}=0 \,, \quad t \geq 0 \, , \quad i,j=1,..,3
\label{eq3}
\end{equation}
where $R_i(t)$ and $v_j(t)$ are the entries of ${\bf R}(t)$ and ${\bf v}(t)$,
respectively,
which  permits to  derive FD1k directly from eqs. (\ref{eq1})-(\ref{eq2}), and 
ii) the equipartition relation
\begin{equation}
\langle v_i \, v_j  \rangle_{\rm eq}= \frac{k_B \, T}{m} \, \delta_{i,j}
\label{eq4}
\end{equation}
The latter relation,
stemming from equilibrium statistical
mechanics  and specifically from the properties of the canonical ensemble,
sets the initial condition for the
velocity autocorrelation function. In eqs. (\ref{eq3})-(\ref{eq4}),
$\langle \cdot \rangle_{\rm eq}$ corresponds to the expected value with
respect to velocity and thermal fluctuations at equilibrium.

Eq. (\ref{eq2}), as well as the  Einstein and Kubo theory, refer 
to particle motion in a unbounded fluid (free space) and in extremely dilute
conditions, in which translational symmetry (homogeneity) applies \cite{kubo2}.

To the Kubo theory applies the observation addressed in \cite{gpp}, regarding
the conditions of dissipative stability and stochastic realizability that
ensure the physical well-posedness of eqs .(\ref{eq1})-(\ref{eq2}).
Indeed, the latter property (stochastic realizability) hinges for the
explicit representation of the fluctuational force ${\bf R}(t)$, and
this can be referred to as the fluctuation-dissipation relation of the
third kind (FD3k). Within the Kubo theory, FD3k automatically
implies FD1k and FD2k. \\

\vspace{0.2cm}
\noindent
{\bf Scope of the present work - }  The scope
of this work is to  address the  generalization  of  FD3k to
particle motion  in generic complex fluids,
in  the case the hydromechanic interactions are non
uniform, i.e.  they  depend on the particle position.
  Physically,
this corresponds either to the  motion  in confined geometries or
to the case of non-diluted suspensions \cite{confined1,confined2,confined3,procgiona_fluid1,suspension1,suspension2}.

This is not only interesting for its physical implications in
microfluidics, suspension rheology, active matter physics, but mainly
because the interplay between spatial nonuniformity and memory
effects in the hydromechanic
response  determines 
 a completely different  and new conceptual problem associated 
with the lack of  knowledge of the  functional expression
for the mean-field
hydrodynamic force that in principle should be derived
from the hydrodynamic field equations.

In order to appreciate this  important issue, 
consider the  confined
motion in a Newtonian fluid, assuming  the  Stokes
regime.  In this case, for a fixed particle position ${\bf x}$,
the dissipative force can be determined by solving
the Stokes equation, thus obtaining 
${\bf F}_{f \rightarrow p}[{\bf v}(t);{\bf x}]=- \boldsymbol{\eta}({\bf x}) \, {\bf v}(t)$,  and thus
 the hydromechanics is characterized by the
 instantaneous friction tensor  $\boldsymbol{\eta}({\bf x})$ that
is positive definite and symmetric.
Since the  response of the fluid is instantaneous, one can argue
that, if the particle position at time $t$ is ${\bf x}(t)$, then
the force acting on the particle at the same time instant is  ${\bf F}_{f \rightarrow p}[{\bf v}(t)]=- \boldsymbol{\eta}({\bf x}(t)) \, {\bf v}(t)$.
Therefore, considering also the action of
an external potential $U({\bf x})$, FD3k is expressed by to the nonlinear Langevin
equation
\begin{eqnarray}
m \, \frac{d {\bf v}(t)}{d t} &= & - \boldsymbol{\eta}({\bf x}(t)) \,
{\bf  v}({\bf x}(t)) - \nabla U({\bf x}(t))  
+  \sqrt{2 \, k_B \, T} \boldsymbol{\eta}^{1/2}({\bf x}(t)) \, \boldsymbol{\xi}(t)  \, , \qquad
\frac{d {\bf x}(t)}{d t}  =  {\bf v}({\bf x}(t))
\label{eq5} 
\end{eqnarray}
where $\boldsymbol{\xi}(t)$ is a 3-d vector of distributional derivatives
of Wiener processes, and $\boldsymbol{\eta}^{1/2}({\bf x})$ 
 the unique symmetric
square root of the friction tensor.
If the potential $U({\bf x})$ enables the establishment
of equilibrium conditions, then: i) the velocity density function is Gaussian and satisfies
the equipartition relation eq. (\ref{eq4}), ii) position ${\bf x}$ and
velocity variables are uncorrelated, and the  marginal  position density
is expressed by 
a Boltzmann distribution,  iii) in the absence of a potential, $U=0$,
in a closed system $\Omega$, the marginal spatial
 density is uniform $p^*({\bf x})=1/\mbox{meas}(\Omega)$,
where $\mbox{meas}(\Omega)$ is the measure of $\Omega$.

Next, consider the motion of the particle in a linear generalized
time-dependent  hydrodynamic regime (here, generalized means that
generic linear constitutive equations for the shear stresses
are considered, such as those arising from linear viscoelasticity).
The only way the hydromechanic fluid-particle interactions can be
determined and expressed analytically in terms of the particle
position and velocity variables, is by solving the hydrodynamic field equations
parametrically with respect to the particle position, customarily
adopting the Laplace transform method \cite{fd1,fd2,fd3,fd4}.
In this way, for a fixed  ${\bf x}$, the Laplace
transform of  
${\widehat {\bf F}}_{f \rightarrow p}[\widehat{\bf v}(s);{\bf x}]= L \left [ {\bf F}_{f \rightarrow p}[{\bf v}(t);{\bf x}]
\right ] =- {\widehat 
{\bf h}}(s;{\bf x}) \, \widehat{\bf v}(s)$
of the force exerted by the fluid onto the particle is obtained, where $s$ is
the Laplace variable  and $\widehat{\bf v}(s)=L[{\bf v}(t)]$.

Observe that
$\widehat{\bf h}(s;{\bf x})$  is the Laplace transform of the memory kernel
for a fixed ${\bf x}$, in which  the particle position plays the
role of a parameter, and  ${\bf h}(t;{\bf x})=L^{-1}[
\widehat{\bf h}(s;{\bf x})]$ is the corresponding kernel in time domain. 

It can be argued,  that the occurrence of confinement and memory effects leads
to generalize eq. (\ref{eq1}) in the form
\begin{equation}
m \, \frac{d {\bf v}(t)}{d t}= - \int_0^t {\bf h}(t-\tau, \{ {\bf x}(\theta)
\}_t )\, {\bf v}(\tau) \, d \tau + {\bf R}(t, \{ {\bf x}(\theta)
\}_t)
\label{eq6}
\end{equation}
where both the kernel ${\bf h}(t-\tau, \{ {\bf x}(\theta)
\}_{t})$, and the fluctuational force ${\bf R}(t, \{ {\bf x}(\theta)
\}_{t})$  are functionals of the particle trajectory $\{x(\theta)\}_t$
from $\theta=0$ up to time $\theta=t$
(due to causality). Nonetheless, the explicit functional dependence
of ${\bf h}(t-\tau, \{ {\bf x}(\theta)
\}_t )$ on the history of particle motion cannot
be recoved from the hydrodynamic analysis performed parametrically
with respect to the particle position, for the simple
reason that ${\bf h}(t-\tau, \{ {\bf x}(\theta)
\}_{t})$ is not equal to ${\bf h}(t;{\bf x})|_{{\bf x}={\bf x}(t)}$,
as arbitrarily stated in \cite{felderhof},
and we have no analytical way to determine, from mechanical reasoning,
 how the memory
kernel entering eq. (\ref{eq6}) would depend on the history
of the particle trajectory. In other words,  
we are in the case where  the fluctuation-dissipation
 theory for a Brownian particle dynamics
should be established in the absence of an exact knowledge of
the mean-field hydrodynamic force.

The astounding and remarkable result derived in this article is that, by enforcing a sound
representation for the hydromechanic interactions and  the very basic properties
of thermodynamic equilibrium, the fluctuation-dissipation analysis
provides a complete solution   both to the  hydrodynamic
problem (explicit representation of the force as a functional of the particle
trajectory history) and to FDk3 namely the explicit representation 
of the thermal
force ${\bf R}(t, \{ {\bf x}(\theta)\}_t)$. In other words,
a compact and unique representation of the mean-field  force elegantly
emerges
from thermodynamic principles.\\

\vspace{0.2cm}

\section{Basic principles}  The complete solution of the
thermal hydromechanic problem (i.e., the explicit expression for 
the mean-field kernel ${\bf h}(t, \{ {\bf x}(\theta)
\}_t )$ and for the thermal force ${\bf R}(t, \{ {\bf x}(\theta)
\}_t)$  entering eq. (\ref{eq6}) in a generic fluid at constant temperature $T$
can be obtained by enforcing three basic principles of general validity. Specifically,
\begin{itemize}
\item the principle of LOCAL REALIZABILITY;
\item the principle of LOCAL CONSISTENCY;
\item the principle of SPATIAL UNIFORMITY at EQUILIBRIUM.
\end{itemize}
Consider the case of a linear viscoelastic fluid, neglecting the effects
of fluid inertia \cite{landau,hydro1,hydro2}. The introduction of these effects does not change the main results, but makes the analysis much more elaborated,  without adding
significant new concepts. It is therefore postponed to a forthcoming
work.

The principle of local realizability, introduced in \cite{gpp}, states that
the force ${\bf F}_{f \rightarrow p}[{\bf v}(t);{\bf x}]$ exerted  by the fluid onto the particle
at constant ${\bf x}$ (i.e. when the particle position ${\bf x}$ is taken as a parameter),
 can be expressed as a linear function of a system
of auxiliary variables 
${\bf z}_1(t),\dots,{\bf z}_N(t)$
the dynamics of which
fulfils a linear system of ordinary differential equations forced at time
$t$ by the local value of the velocity ${\bf v}(t)$.
Following \cite{gpp}, this essentially
 implies that the force could be expressed as
\begin{equation}
{\bf F}_{f \rightarrow p}[{\bf v}(t);{\bf x}]= -\sum_{i=1}^N
{\bf A}_i({\bf x}) \, \lambda_i \, e^{-\lambda_i  t} * {\bf  v}(t)
\label{eq7}
\end{equation}
where ${\bf A}_i({\bf x})$ are $3 \times 3$ matrices depending on ${\bf x}$,
$\lambda_i$ are (scalar and position independent) relaxation rates, and ``$*$'' indicates
convolution. The reason for this is simple. Introducing the auxiliary memory
variables ${\bf z}_i(t)= e^{-\lambda_i  t} * {\bf  v}(t)$, particle dynamics reduces to a stationary
system of linear differential equations in $({\bf v},{\bf z}_1,\dots,{\bf z}_N$).

This form of memory kernel, in which the relaxation rates are position
independent, emerges either from rheology and from
hydrodynamics (fluid inertial effect). From hydrodynamics in confined
geometries (e.g. considering a Maxwell fluid, or a viscoelastic fluid
characterized by $N$ relaxation rates) \cite{franosch,viscoelastic1,viscoelastic2}, the $\lambda_i$'s can be
realistically assumed to be position independent, so that
nonuniformities induced by confinement  enter 
 exclusively in the position-dependent matrices ${\bf A}_i({\bf x})$.

The second principle, namely the principle of local
consistency permits to define the functional representation  either for  the hydromechanic
or  the
fluctuational force. The principle can be stated as follow:
consider the exact formulation of the equations of motion
of a Brownian particle, in which the particle position ${\bf x}(t)$
is a dynamic variables changing in time according to the kinematic
equation. If we consider the dynamic equations for ${\bf v}(t)$
and $({\bf z}_1(t), \times {\bf z}_N)$ in the case particle position
is kept constant, say ${\bf x}={\bf x}^*$, these equations
should correspond to that of an effective Brownian particle
possessing a memory kernel ${\bf h}_{\rm eff}(t)={\bf h}(t;{\bf x}^*)$
and an effective thermal force ${\bf R}_{\rm eff}(t)={\bf R}(t;{\bf x}^*)$,
and thus classical fluctuation-dissipation theory can be applied to
these idealized conditions. In the essence, the local consistency
limits the spectrum of admissible hydromechanics models for particle
dynamics.

Indeed, if we  gather these two principles  (for details
see Appendices \ref{appA} and \ref{appB}) we can introduce $2 \, N$
symmetric and positive definite
 matrices $\boldsymbol{\alpha}_i({\bf x})$, $\boldsymbol{\beta}_i({\bf x})$, $i=1,\dots,N$, commuting
with each other, 
related to the hydromechanic matrices entering eq. (\ref{eq7}) via
the conditions
\begin{equation}
\boldsymbol{\alpha}_i({\bf x}) \, \boldsymbol{\beta}_i({\bf x}) = {\bf A}_i({\bf x}) \,,
\quad i=1,\dots,N
\label{eq8}
\end{equation} 
such that  the local representation of the particle dynamics
expressed in terms 
of  the $N$ internal degrees of
freedom ${\bf z}_i(t)$, $i=1,\dots,N$,  
takes the functional form
\begin{eqnarray}
\frac{d {\bf x}(t)}{d t} & = & {\bf v}(t) \nonumber \\
m \frac{d {\bf v}(t)}{d t} & = & - \sum_{i=1}^N \boldsymbol{\alpha}_i({\bf x}(t)) \, \lambda_i \, {\bf z}_i(t) \label{eq9} \\
\frac{d {\bf z}_i(t)}{d t} & = & -\lambda_i \, {\bf z}_i(t) + \boldsymbol{\beta}_i({\bf x}(t)) \,{\bf v}(t)
+ \sqrt{2} \, {\bf c}_i({\bf x}(t)) \, \boldsymbol{\xi}_i(t)
\nonumber
\end{eqnarray}
where $\boldsymbol{\xi}(t)$, $i=1,\dots,N$ are  the distributional
derivatives of $N$ independent  3-d  Wiener processes.

Owing to  eqs. (\ref{eq8}), the principle of local consistency is satisfied for any ${\bf x}(t)={\bf x}^*$,
and
the coefficients
${\bf c}_i({\bf x}^*)$ can be determined enforcing classical
Kubo fluctuation-dissipation theory, i.e. the validity of FD1k and FD2k, at constant ${\bf x}$ \cite{goychuk},
and this leads to
the expression
\begin{equation}
{\bf c}_i({\bf x})= \sqrt{k_B \, T} \, \left (\boldsymbol{\alpha}_i^{-1}({\bf x}) \boldsymbol{\beta}_i({\bf x}) \right )^{1/2}
\label{eq10}
\end{equation}
$i=1,\dots,N$.\\

\vspace{0.2cm}
\noindent
\section{Main results} There are infinitely many  admissible
representations eq. (\ref{eq9}) of particle dynamics, as the
matrices $\boldsymbol{\alpha}_i({\bf x})$ and $\boldsymbol{\beta}_i({\bf x})$ are arbitrary,
apart from the condition eq. (\ref{eq8}). To define these matrices  uniquely, thermodynamic
properties should be enforced. Specifically, it is sufficient to enforce the very basic 
property of  equilibrium conditions, namely the absence of spatial gradient.
This can be state as follows. Consider the motion of a particle subjected to
eq. (\ref{eq10}) in a closed  and bounded domain $\Omega$ at constant temperature $T$ and in
the absence of external forces acting on it. At equilibrium the marginal spatial
density function should be uniform.  This
is the meaning of the very general principle of spatial
uniformity at equilibrium. We have the following main results:\\
\subsection*{ Theorem   I } There exists a unique, position independent equilibrium
solution $p_{\rm eq}({\bf x},{\bf v},\{{\bf z}_i\}_{i=1}^N)$ 
of the Fokker-Planck equation associated with eq. (\ref{eq9})
if 
\begin{equation}
\boldsymbol{\alpha}_i({\bf x})= \boldsymbol{\beta}_i({\bf x})= {\bf A}_i^{1/2}({\bf x})
\label{eq11}
\end{equation}
In this case   
\begin{equation}
 p_{\rm eq}({\bf x},{\bf v},\{{\bf z}_i\}_{i=1}^N) = C \, \mbox{exp} \left [
- \frac{m |{\bf v}|^2}{2 \, k_B \, T} - \sum_{i=1}^N \frac{\lambda_i \, |{\bf z}_i|^2}{2 \, k_B \, T}
\right ]
\label{eq12}
\end{equation}
where $C$ is the normalization constant $\diamond$.\\

 The proof of this and of the forthcoming proposition can be found in Appendix \ref{appB}.
It follows from this result an important implication.\\
\subsection*{Theorem II} For the hydromechanic problem expressed by eq. (\ref{eq9})
the thermodynamically consistent expression for the force exerted by the fluid on the
particle is
\begin{eqnarray}
{\bf F}_{f \rightarrow p}[{\bf v}(t)] & = &
\sum_{i=1}^N {\bf A}_i^{1/2}({\bf x}(t)) \, \lambda_i \,\int_0^t e^{-\lambda_i (t-\tau)} 
 \, {\bf A}_i^{1/2}({\bf x}(\tau)) \, {\bf v}(\tau) \, d \tau
\label{eq13}
\end{eqnarray}
and the overall thermal force attains the expression 
\begin{equation}
{\bf R}(t,{\bf x }(t)) = \sqrt{2 \, k_B \, T} \sum_{i=1}^N {\bf A}_i({\bf x}(t)) \, \lambda_i
\, \int_0^t e^{-\lambda_i (t-\tau)}
\, \boldsymbol{\xi}_i(\tau) \, d \tau
\label{eq14}
\end{equation}
and it depends solely on the actual value of particle position ${\bf x}(t)$ $\diamond$.\\

Moreover, if we further assume a more stringent definition
of thermodynamic equilibrium, in the meaning that all the
statistical properties of the particle velocity ${\bf v}$ and
of the auxiliary variables ${\bf z}_1,\dots,{\bf z}_N$ should
be position independent, the condition stated in Theorem I is not only
sufficient but also necessary.

These two propositions completes the analysis of the problem.
The other case of memory effects of hydrodynamic relevance, namely
the inclusion of fluid-intertial memory effects (determining in the
Newtonian case the occurrence of the Basset force) \cite{landau,hydro1,hydro2}
can be conceptually treated in the same way, by
further considering that the finite propagation of the shear stresses associated
with viscoelastic effects determines
the non singularity of the fluid-inertial memory kernel at $t=0$ \cite{pg_visco}.
The technical and notational set-up of this case is rather delicate and lengthy,
not suitable for the goals of a short communication, and it 
will be thoroughly developed in a subsequent work.

Observe that both FD3k and the hydromechanical problem of particle motion are completely
solved by eqs. (\ref{eq13}) and (\ref{eq14}). Conversely, no explicit analytical
expressions can be provided for the autocorrelation function of the particle
velocity and of the thermal force (corresponding to FD1k and FD2k) due to the intrinsic
nonlinear dependence  on the particle position ${\bf x}(t)$ of the hydromechanical
matrices entering eqs. (\ref{eq9})-(\ref{eq11}). In any case, these
functions can be estimated directly from the stochastic
simulations of eq. (\ref{eq9}). 

Another interesting observation involves the functional dependence  of the thermal
force on the position. From eqs.  (\ref{eq10})-(\ref{eq11}) we have that the
coefficient matrices ${\bf c}_i$ are position independent in the thermodynamically
consistent case, ${\bf c}_i= \sqrt{k_B \, T} \, {\bf I}$, where ${\bf I}$ is the identity matrix,
and the resulting overall thermal force ${\bf R}(t,{\bf x}(t))$  at time $t$ depends solely
on the actual position ${\bf x}(t)$ and not on the kinematic history of the
particle.  Adopting the widely accepted classification of stochastic
differential equations, this means that eqs. (\ref{eq9})-(\ref{eq11}) 
represent a system
of linear Langevin equation, as the amplitudes
 of the stochastic forcings do not depend on
the particle state coordinates. Conversely, the mean-field hydrodynamic  force eq. (\ref{eq13})
admits a ``symmetric'' dependence on the actual particle position at time $t$ and on the
kinematic hystory of the particle, stemming from the equality of the
$\boldsymbol{\alpha}_i({\bf x})$ and $\boldsymbol{\beta}_i({\bf x})$
expressed by eq. (\ref{eq11}).  Finally, observe that the theory
 would not have been possible
develop without the extensive use of the representation
underlying the principle of local realizability, as it clear emerges
from the functional form of the thermodynamically consistent
mean-field force eq. (\ref{eq13}).\\

\vspace{0.2cm}
\noindent
\section{ Examples and applications } To make a simple example,
 consider the  motion of a spherical particle of radius $R_p$ and mass $m$
between
two parallel plates at distance $W+2 \, R_p$, and let $x \in [0,W]$ be the particle distance
from one of the plates,  
in the presence of a simple Maxwell fluids characterized
by a relaxation rate $\lambda$.
In this case, considering solely the motion orthogonal to the walls, 
 the memory kernel is $h(t,x)= \eta_0 \, \widetilde{\eta}
(x) \, \lambda e^{-\lambda  t}$,   where 
$\widetilde{\eta}(x)$ accounts for the hydromechanical effects due to
the confinement.
In dimensionless form, still letting $x$, $v$, and $z$ be the dimensionless
position, velocity and auxiliary variable accounting for the Maxwell model,
with $x \in (0,1)$,
the equations of motion eq. (\ref{eq9})  read
\begin{eqnarray}
\frac{d x(t)}{d t} & = & v \nonumber \\
\frac{d v(t)}{d t} & = & - \zeta \, \alpha(x(t)) \, z(t) 
\label{eq15}
\\
\frac{d z(t)}{d t} & = & -\gamma\, z(t) + \zeta \, \beta(x(t)) \, v(t) + \sqrt{\frac{2 \, \beta(x(t))}{\alpha(x(t))}} \, \varepsilon \, \xi(t)
\nonumber
\end{eqnarray}
where $\zeta$, $\gamma$, and $\varepsilon$ are constant
depending on two nondimensional parameters $\widetilde{\lambda}$
and $\ell_c$, $\zeta=\sqrt{\widetilde{\lambda}} \, \ell_c$, $\gamma=\widetilde{\lambda} \, \ell_c$, and $\varepsilon=\sqrt{\gamma}$,
and $\alpha(x)$, $\beta(x)$ are the hydromechanic factors.
We consider the general case, i.e.,
  $\alpha(x)= A^p(x)$, $\beta(x)=A^{1-p}(x)$ with $p \in [0,1]$, where $A(x)$
is  the nondimensional representation of $\widetilde{\eta}(x)$  in the mean-field  approximation
\begin{equation}
A(x)= \frac{(c+x) \, (c+1-x)}{(x) \, (1 -x)}
\label{eq16}
\end{equation}
with $0<c<1$, where $p=1/2$  is the thermodynamically correct value.
The above nondimensional  representation
provides $\langle v^2 \rangle_{\rm eq}=\langle z^2 \rangle_{\rm eq}=1$
in the thermodynamically consistent case ($p=1/2$). 
 For  details see Appendix \ref{appC}.  An ensemble of $N_p=10^5$
particles initially uniformly distributed in $x \in (0,1)$, and with
$v(0)=z(0)=0$ has been considered. Reflective boundary conditions
are enforced at $x=0,1$.

\begin{figure}
\includegraphics[width=6cm]{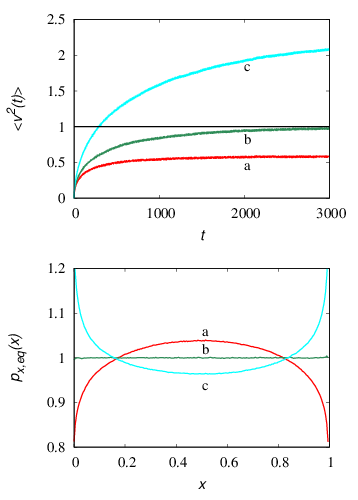}
\caption{ (upper panel) $\langle v^2(t) \rangle$ vs $t$.  (lower panel)
Equilibrium marginal spatial density $p_{x, {\rm eq}}(x)$ vs $x$.
The data refers to $\ell_c=1$, $\widetilde{\lambda}=0.1$, $c=0.1$.
Curves (a) represent $p=1$, (b) $p=1/2$, (c) $p=0$.}
\label{Fig1}
\end{figure}

Figure \ref{Fig1} depicts the dynamics of $\langle v^2(t) \rangle$
for  three values of $p$. As expected, for $p=1/2$, the
mean squared velocity reaches the equilibrium value of $1$,  and lower/higher values
of the velocity variance occur for $p>1/2$ and for $p<1/2$, respectively. This
deviation from the correct equilibrium value is the consequence of a nonuniform
spatial particle distribution, as can be evinced from the analysis of the
marginal spatial density function at equilibrium $p_{x,{\rm eq}}(x)= \int 
\int p_{\rm eq}(x,v,z) \, d v \, dz$,
depicted in  the lower panel of figure \ref{Fig1}. Data refer to an equilibrium ensemble of $10^8$ realizations. This results confirms the theory. 

Due to its generality, the theory can be extended in a manifold of different applications. It provides
a way to develop thermodynamically consistent models of suspensions in which the
equations of motion may depend either on the relative hydrodynamic friction amongst particles
(as in Brady and Bossis Stokesian dynamics) \cite{brady1,brady2} or on the particle marginal spatial densities (as in
non-linear stochastic models interpreted a la McKean) \cite{mckean1,mckean2}. 
In the former case (i.e. Stokesian
dynamics), the present theory provides a way to extend this approach to include
fluid-inertial effects, which, as known, control short-time properties and 
the shape of the velocity autocorrelation function \cite{expo0,expo1,expo2,expo3}.

Another  interesting result involves active particles, i.e. the  class of problems
in which the particle is subjected to additional stochastic fluctuations superimposed
to the thermal force \cite{active1,active2}. Let us make a simple example. Consider the dynamics
of active particles in a  Maxwell-fluid.
This corresponds to  eq. (\ref{eq15}) in which the second equation is
replaced by
\begin{equation}
\frac{d v(t)}{d t}  =  - \zeta \, \alpha(x(t)) \, z(t) +\sqrt{2} \, \varepsilon_a \, \xi_1(t)
\label{eq17}
\end{equation}
where
$\varepsilon_a$ represents the intensity of the active fluctuations
modeled as a distributional derivative of a Wiener process $\xi_1(t)$ independent
of $\xi(t)$. 
In this case assume $\alpha(x)=A^{1/2}(x)$ as
dictated by  thermodynamic consistency. This is just an example, and other class of active fluctuations
can be considered as well.
First of all,  the presence of an additional active
fluctuation source determines the violation of the classical equilibrium distribution,
in which $v$, $z$ and the position $x$ are independent stochastic variables.
This is addressed in Appendix \ref{appD}. As a consequence, anomalies in
equilibrium properties may occur.

\begin{figure}
\includegraphics[width=6cm]{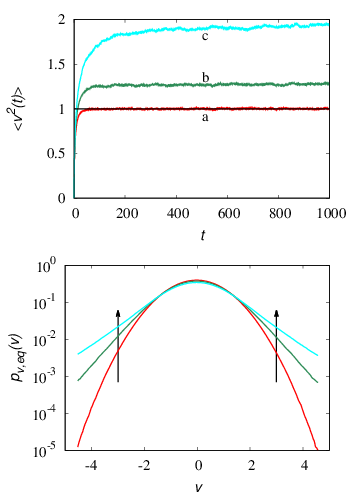}
\caption{$\langle v^2(t) \rangle$ vs $t$ (upper panel), and equilibrium
velocity density $p_{v,{\rm eq}}(v)$ vs $v$ (lower panel) for
the active particle model eqs. (\ref{eq15})-(\ref{eq17}) at $\ell_c=1$,
$\widetilde{\lambda}=1$, $c=0.1$, for different values of $\varepsilon_a$ (the arrows indicate increasing values of $\varepsilon_a$: (a) $\varepsilon_a=0$,
(b) $\varepsilon_a=0.1$, (c) $\varepsilon_a=0.15$.}
\label{Fig2}
\end{figure}

Figure \ref{Fig2} (upper panel) depicts the evolution of $\langle v^2(t) \rangle$ for different
values of  $\varepsilon_a$, starting from quiescent conditions. It is sufficient a small amount of external (active) randomness to
determine a significant increase of $\langle v^2 \rangle$ beyond the equilibrium value
(obtained for $\varepsilon_a=0$). Moreover, 
even small values of $\varepsilon_a$ determine a deviation for the
Maxwellian equilibrium distribution at steady state, occurring in
the passive case $\varepsilon_a=0$. This is depicted in figure \ref{Fig2} (lower panel)
showing the  marginal steady-state  distribution $
p_{v,{\rm eq}}(v)=\int \int p_{\rm eq}(x,v,z) \, dx \, d z$.
This   and other problems in confined systems  will be addressed
more thoroughly in forthcoming works. \\

\newpage

\vspace{0.2cm} 
\noindent
{\bf Acknowledgment - } This research  received financial support from ICSC---Centro Nazionale di Ricerca in High Performance Computing, Big Data and Quantum Computing, funded by European Union---NextGenerationEU.

\appendix
\section{Discussion on basic principles}
\label{appA}
It is useful to address
 in greater detail the principles stated in the manuscript, upon which the 
generalized fluctuation-dissipation theory in
confined geometries is built upon.
\begin{itemize}
\item {\bf The principle of LOCAL REALIZABILITY - } It involves exclusively
the mean-field hydromechanic  contribution to fluid-particle interactions.
The principle states that the  memory integral in
the expression for ${\bf F}_{f \rightarrow p}[{\bf v}(t)]$ at time $t$ is
just a compact representation of an underlying local, linear and stationary dynamics
involving additional  auxiliary variables, say ${\bf z}_1(t),\dots,{\bf z}_N(t)$,
and such that, in this extended representation of  fluid-particle interactions,
the dynamics of the auxiliary variables depend  exclusively on
the local value of ${\bf v}(t)$ at the actual time $t$.

In order to clarify further this issue, consider the case of an unbounded fluid
(no confinement) for which
\begin{equation}
{\bf F}_{f \rightarrow p}[{\bf v}(t)] = - \int_0^t {\bf h}(t-\tau) \, {\bf v}(\tau) \, d \tau
\label{eq01}
\end{equation}
and neglect the contribution of thermal fluctuations that is immaterial
in the present analysis.

For the  force 
to admit a local linear representation,  ${\bf F}_{f \rightarrow p}[{\bf v}(t)]$ should be expressed as a linear function of some
additional variables ${\bf z}_1(t),\dots,{\bf z}_N(t)$, (where $N$ is either
finite or countable). This means that there exist $N$ constant-coefficient
$3 \times 3$ matrices ${\bf B}_i$, $i=1,\dots,N$, such that
\begin{equation}
{\bf F}_{f \rightarrow p}[{\bf v}(t)]  = - \sum_{i=1}^N {\bf B}_i \, {\bf z}_i(t)
\label{eq02}
\end{equation}
and the variables $\{ {\bf z}_i(t) \}_{i=1}^N$
 satisfy a linear system of differential equations forced by the actual value of the
particle velocity. Without loss of generality, we may assume a decoupling
between the ${\bf z}_i(t)$ (although this assumption could be easily removed, if necessary),
and express the internal dynamics of the auxiliary variables
in the form
\begin{equation}
\frac{d {\bf z}_i(t)}{d t}=-\lambda_i \, {\bf z}_i(t) + {\bf v}(t) \, ,
\qquad i=1,\dots,N
\label{eq03}
\end{equation}
where $\lambda_i>0$ are the  relaxation rates of the internal modes.
Eq. (\ref{eq03}) represents a system of linear differential equations
with constant coefficients depending 
exclusively on the actual value ${\bf v}(t)$ of the particle
velocity.
Integrating eq. (\ref{eq03}), we have
\begin{equation}
{\bf z}_i(t)= e^{-\lambda_i  t} \, {\bf z}_i(0) + \int_0^t e^{-\lambda_i  (t-\tau) } \, {\bf v}(\tau) \, d \tau= e^{-\lambda_i  t} \, {\bf z}_i(0) + 
e^{-\lambda_i t} * {\bf v}(t)
\label{eq04}
\end{equation}
where the symbol ``$*$'' indicates convolution.
The first term involves the initial condition for the internal variables,
decaying exponenially to zero, and therefore  it is immaterial in the long-term. So
we can always set ${\bf z}_i(0)=0$. In this way,  the
compact representation for  ${\bf F}_{f \rightarrow p}[{\bf v}(t)] $ follows
\begin{equation}
{\bf F}_{f \rightarrow p}[{\bf v}(t)] = - \sum_{i=1}^N {\bf B}_i \, e^{-\lambda_i \, t} * {\bf v}(t) 
\label{eq05}
\end{equation}
We can always introduce a new system of matrices ${\bf A}_i$, such that
${\bf B}_i={\bf A}_i \, \lambda_i$ so that eq. (\ref{eq05}) becomes
\begin{equation}
{\bf F}_{f \rightarrow p}[{\bf v}(t)] = - \sum_{i=1}^N {\bf A}_i \, 
\lambda_i \, e^{-\lambda_i
\, t} * {\bf v}(t) 
\label{eq06}
\end{equation}
corresponding to the expression used in the manuscript.

In practice, the principle of local realizability claims the
existence of a Markovian embedding for the mean-field hydromechanic interactions. Whether this statement should be viewed as a principle or more simply
 as a useful
approximation/representation that finds application in all the cases of physical interest, is a questionable matter of  formal sensitivity.
Nevertheless, we stress its conceptual importance (as further addressed
in the case of confined systems), rising it to
the level of a principle, in order to underline
the physical constraint of temporal locality of interactions
in classical physics,
 that dictates that the only way memory effects may arise in physical systems is due to the background evolution of internal degrees of freedom, characterized by
local dynamics. In the context of fluid-particle interactions,
the internal degrees of freedom are associated with the
family of hydrodynamic modes characterizing the hydrodynamic
problem and copying with the particle at the level of a boundary condition.

\item {\bf The principle of LOCAL CONSISTENCY - }  From one hand, this
principle matches the way hydromechanic fluid-particle interactions 
are  evaluated,  from the other hand it represents a conceptually
evident consistency principle. In any case, it provides the front door to
address fluctuation-dissipation relations in confined systems.

In confined linear hydrodynamic  problems, such as those of interest
in the present case, the expression for 
${\bf F}_{f \rightarrow p}[{\bf v }(t)]$ cannot be directly derived by
solving the linear hydrodynamic field equations equipped with
boundary conditions at the walls of the confinement and at the particle
external surface. What can
be achieved, and it is customarily developed in fluid-dynamic
literature, is to  tranform the linear hydrodynamic equations in the Laplace (or Fourier)  domain to obtain, either analytically or numerically,
  the Laplace transform  $\widehat{\bf F}_{f \rightarrow p}[\widehat{\bf v }(s);{\bf x}]$ of 
${\bf F}_{f \rightarrow p}[{\bf v }(t);{\bf x}]$ at a constant value  ${\bf x}$
of  the particle position.  Linearity and stationarity dictate
that  $\widehat{\bf F}_{f \rightarrow p}[\widehat{\bf v }(s);{\bf x}]$ should
be of the form 
\begin{equation}
\widehat{\bf F}_{f \rightarrow p}[\widehat{\bf v}(s);{\bf x}]=
-\widehat{\bf h}(s;{\bf x}) \, \widehat{\bf v}(s)
\label{eq07}
\end{equation}
where hatted-symbols corresponds to the Laplace transforms of
the corresponding quantities functions of time,  say $\widehat{\bf v}(s)=L[{\bf v}(t)]$, and $s$ is the
Laplace variable.
Thus, by the principle of local realizability we have in time domain
\begin{equation}
{\bf h}(t;{\bf x})= L^{-1} \left [ \widehat{\bf h}(s;{\bf x}) \right ] = \sum_{i=1}^N {\bf A}_i({\bf x}) \, \lambda_i
\,  e^{-\lambda_i  t}
\label{eq08}
\end{equation}
where the $3 \times 3$ matrices ${\bf A}_i({\bf x})$, $i=1,\dots,N$,
depend parametrically on the particle position ${\bf x}$.

In the unbounded case ${\bf A}_i({\bf x})={\bf A}_i$ are constant-coefficient matrices, and eq. (\ref{eq08}) provides a complete description of the mean-field
hydromechanics. Conversely, due to the explicit parametric dependence
on ${\bf x}$ characterizing confined problems, eq. (\ref{eq08})
does not solve, in this more general case,
 the hydromechanics of fluid-particle interactions,
 as  it is not possible to disclose from this
expression the proper and correct dependence of the mean-field force 
on  the history of particle trajectory, as ${\bf x}(t)$
is a function of time and not a mere parameter.

Nevertheless, this parametric setting can still be used in a gedanke
experiment, in which  the dynamic equations of motion,
and thus the resulting fluctuation-dissipation relations, are considered
at constant ${\bf x}$, i.e. in the situation in which the
particle position is just a parameter.

This conceptual setting is less unrealistic than it may seem at a
first sight. For instance,
one may consider the axial motion of spherical particles in straight channels, in
which the distance of the particle from the channel walls remains constant.
Since all the hydromechanics matrices ${\bf A}_i({\bf x})$ depend
in this configuration on the particle distance from the
channel walls, they remain constant in this particular case,
and therefore the particle motion, as regards the dynamic equations
(i.e., neglecting particle kinematics) is equivalent to
the configuration at which the particle position remains
fixed and equal to its initial value, say ${\bf x}^*$.
In this case, the effective hydromechanic kernel
is given by
\begin{equation}
{\bf h}_{\rm eff}(t)={\bf h}(t;{\bf x}^*)
\label{eq09}
\end{equation}
where the quantity at the r.h.s. of eq. (\ref{eq09}) 
is given by eq. (\ref{eq08}) evaluated
at ${\bf x}={\bf x}^*$.
Consequently, the dynamics of Brownian motion is fully specified
by the equation
\begin{equation}
m \,  \frac{d {\bf v}(t)}{d t}= - {\bf h}_{\rm eff}(t) * {\bf v}(t)
+ {\bf R}_{\rm eff}(t)
\label{eq010}
\end{equation}
where the effective termal force ${\bf R}_{\rm eff}(t)$ is
given 
\begin{equation}
{\bf R}_{\rm eff}(t)={\bf R}(t;{\bf x}^*)
\label{eq011}
\end{equation}
and, for any fixed ${\bf x}^*$  eq. (\ref{eq010})  is
equivalent to a Brownian particle problem in the unconfined
case, in which ${\bf R}_{\rm eff}(t)$, and thus ${\bf R}(t;{\bf x}^*)$
can be evaluated as in the classical Kubo theory.

The principle
of local consistency states that if the correct equation
of motion of the Brownian particle in a confined situation
is 
\begin{equation}
m  \, \frac{d {\bf v}(t)}{d t}= - \int_0^t {\bf h}(t-\tau, \{ {\bf x}(\theta)
\}_t )\, {\bf v}(\tau) \, d \tau + {\bf R}(t, \{ {\bf x}(\theta)
\}_t)
\label{eq012}
\end{equation}
then, in the case ${\bf x}$ is kept constant and equal to ${\bf x}^*$,
 it should
reduce to
\begin{equation}
m  \, \frac{d {\bf v}(t)}{d t}= - \int_0^t {\bf h}(t-\tau,  {\bf x}^*)
\, {\bf v}(\tau) \, d \tau + {\bf R}(t,  {\bf x}^*)
\label{eq013}
\end{equation}
In this way, the functional form of ${\bf R}(t,  {\bf x}^*)$
can be evaluated using classical methods, parametrically with respect to
${\bf x}^*$. This analytical calculations
stemmming from this approach are reported in detail in the next paragraph.

The principle of local consistency provides a way for defining
classes of consistent equations of motion.
Enforcing  local representability,
 the equation of motion
of a particle can be expressed
in the general form
\begin{eqnarray}
m \, \frac{d {\bf v}(t)}{d t} & = & - \sum_{i=1}^N   \boldsymbol{\alpha}_i({\bf x}(t)) \, \lambda_i \, {\bf z}_i(t)
\nonumber \\
\frac{d {\bf z}_i(t)}{d t} & = & - \lambda_i \, {\bf z}_i(t) + 
\boldsymbol{\beta}_i({\bf x}(t)) \, {\bf v}(t)+ \sqrt{2} \, {\bf c}_i({\bf x}(t)) \, \boldsymbol{\xi}_i(t) \, , \quad i=1,\dots,N
\label{eq014}
\end{eqnarray}
where $\boldsymbol{\xi}_i(t)=(\xi_{i,1}(t),\xi_{i,2}(t),\xi_{i,3}(t))$
in a vector of distributional derivatives of independent Wiener
processes, $\langle \xi_{i,h}(t) \, \xi_{j,k}(t^\prime) \rangle =
\delta_{ij}\, \delta_{hk} \, \delta(t-t^\prime)$, while the matrices
$\boldsymbol{\alpha}_i({\bf x})$, $\boldsymbol{\beta}_i({\bf x})$
and ${\bf c}_i({\bf x})$  have to be determined.
At this stage we can enforce local consistency.
Consider the case at constant ${\bf x}$. From the second set
of equations (\ref{eq014}), setting ${\bf z}_i(0)=0$, we
have
\begin{equation}
{\bf z}_i(t)= \boldsymbol{\beta}_i({\bf x}) \, e^{-\lambda_i  t} * {\bf v}(t)
+ \sqrt{2} \, {\bf c}_i({\bf x}) \, e^{-\lambda_i  t} * \boldsymbol{\xi}_i(t)
\label{eq015}
\end{equation}
Thus, substituting into the first equation eq. (\ref{eq014}),
\begin{equation}
m \frac{d {\bf v}(t)}{d t} = - \sum_{i=1}^N \boldsymbol{\alpha}_i({\bf x})
\, \boldsymbol{\beta}_i({\bf x}) \, \lambda_i \, e^{-\lambda_i  t} * {\bf v}(t)
- \sqrt{2} \, \sum_{i=1}^N \boldsymbol{\alpha}_i({\bf x}) \,
{\bf c}_i({\bf x})  \, \lambda_i \, e^{-\lambda_i  t} * \boldsymbol{\xi}_i(t)
\label{eq016}
\end{equation}
that compared with eq. (\ref{eq07}) provides
\begin{equation}
\boldsymbol{\alpha}_i({\bf x}) \, \boldsymbol{\beta}_i({\bf x})=
{\bf A}_i({\bf x}) \, , \qquad i=1,\dots, N
\label{eq017}
\end{equation}
that represent the constraints on these matrices deduced
by the principles of local representability and consistency.
Moreover, as  developed in the next paragraph, the expression
for the matrices ${\bf c}_i({\bf x})$ immediately follows by classical
fluctuation-dissipation analysis.

It is rather clear that eqs. (\ref{eq014}) constitute the most
general representation for the particle hydromechanics. Local
representability permits to unveil such a general representation,
decomposing the effect of particle location into the two
families of hydromechanic matrices $\boldsymbol{\alpha}_i({\bf x})$,
$\boldsymbol{\beta_i}({\bf x})$.
Local consistency sets some fundamental constraints, specifically
eq. (\ref{eq017}), on the nature of these matrices, and determines the intensity
of the thermal fluctuations acting on the internal degrees of freedom, i.e.
the expression for the matrices ${\bf c}_i({\bf x})$.

While in the unbounded case, i.e. whenever these matrices are constant
and do not depend on particle position, any choice of the $\boldsymbol{\alpha}_i$ and $\boldsymbol{\beta}_i$ matrices, consistent with eq. (\ref{eq017}),
provides an equivalent representation for particle dynamics,  the choice
of these matrices
becomes crucial in the confined case, as it determines completely different
formulations of particle hydromechanics.
To see this, consider again the second equation (\ref{eq014})
in the  dynamic case where ${\bf x}(t)$ is no longer a parameter
but a function of time evolving according to the kinematic equation. The
${\bf z}_i(t)$ auxiliary variables can be explicited (taking, as
discussed above, ${\bf z}_i(0)=0$),
\begin{equation}
{\bf z}_i(t)= e^{-\lambda_i t} * \left [ \boldsymbol{\beta}_i({\bf x}(t))
\, {\bf v}(t) \right ] + \sqrt{2} e^{-\lambda_i t} *
\left [ {\bf c}_i({\bf x}(t)) \, \boldsymbol{\xi}_i(t) \right ]
\label {eqa1}
\end{equation}
that, substituted into the first  equation (\ref{eq014}), provide
\begin{equation}
m \frac{d {\bf v}(t)}{d t}= - \sum_{i=1}^N \boldsymbol{\alpha}_i({\bf x}(t))
\, \lambda_i \, e^{-\lambda_i t} * \left [ \boldsymbol{\beta}_i({\bf x}(t))
\, {\bf v}(t) \right ] - \sqrt{2} \sum_{i=1}^N \boldsymbol{\alpha}_i({\bf x}(t))
\, \lambda_i \, e^{-\lambda_i t} *
\left [ {\bf c}_i({\bf x}(t)) \, \boldsymbol{\xi}_i(t) \right ]
\label {eqa2}
\end{equation}
from which the expression for the mean-field hydrodynamical
force and for the thermal force follow
\begin{equation}
{\bf F}_{f \rightarrow p}[{\bf v}(t)]= - \sum_{i=1}^N \boldsymbol{\alpha}_i({\bf x}(t))
\, \lambda_i \int_0^t  e^{-\lambda_i (t-\tau)}\,  \boldsymbol{\beta}_i({\bf x}(\tau))
\, {\bf v}(\tau)  \, d \tau
\label{eqa3}
\end{equation}
\begin{equation}
{\bf R}(t,\{{\bf x}(\theta)\}_t)=
 - \sqrt{2} \sum_{i=1}^N \boldsymbol{\alpha}_i({\bf x}(t))
\, \lambda_i \, \int_0^t e^{-\lambda_i (t-\tau)} 
\, {\bf c}_i({\bf x}(\tau)) \, \boldsymbol{\xi}_i(\tau)  \, d \tau
\label {eqa4}
\end{equation}
From the expressions eqs. (\ref{eqa3})-(\ref{eqa4}) it becomes evident the
different role of these matrices. The $\boldsymbol{\alpha}_i({\bf x}(t))$
matrices account for the instataneous influence of the hydromechanic 
constraints at time $t$ acting on the particle, while $\boldsymbol{\beta}_i({\bf x}(t))$ refer to the contribution of the past history of the
particle trajectory, representing the memory modulation of the force 
 deriving from  the
past positions visited by the particle.
As local consistency does not provide a unique representation
of particle hydromechanics, the further addition of thermodynamic
constraints solve univocally the problem, yielding a unique
representation of particle motion, as developed below.

Observe from eq. (\ref{eqa3}) that the functional form of the
force exerted by the fluid onto the particle is no longer of
convolutional nature. Convolutional structures, as a response to
a forcing term, arise in linear stationary (autonomous) systems
due to causality. But in the present case the system is not linear
due to explicit nonlinear dependence of the hydromechanic matrices ${\bf A}_i({\bf x})$ on the position ${\bf x}$ and this explans the structure of
eq. (\ref{eqa3}).

It should be further   stressed the importance of the modal
decomposition deriving from the principle of local realizability, without
which eqs.  (\ref{eqa3})-(\ref{eqa4}) could not be derived.

Let us further discuss eq. (\ref{eq017}).
From eq. (\ref{eq017}), it is reasonable to assume that
there exist two systems of scalar functions $f_i$, $g_i$, $i=1,\dots,N$
such that
\begin{equation}
\boldsymbol{\alpha}_i({\bf x})= f_i({\bf A}_i({\bf x})) \, ,
\qquad \boldsymbol{\beta}_i({\bf x})= g_i({\bf A}_i({\bf x}))
\, , \qquad i=1,\dots,N
\label{eq018}
\end{equation}
with  the property that
\begin{equation}
f_i(x) \, g_i(x) = x \, , \qquad i=1,\dots, N
\label{eq019}
\end{equation}
and this implies that  the matrices $\boldsymbol{\alpha}_i({\bf x})$ and $\boldsymbol{\beta}_i({\bf x})$ commute for any $i=1,\dots,N$.
Finally, the positive definiteness of these matrices ensure the stochastic
realizability of the particle dynamics, i.e. the existence
of the matrices ${\bf c}_i({\bf x})$.

\item {\bf The principle of SPATIAL UNIFORMITY at EQUILIBRIUM - }
This principle expresses the most fundamental property at equilibrium.
This principle states that a Brownian particle immersed in a quiescent fluid 
within a closed and bounded domain at constant temperature $T$,  in the
absence of external forces, potentials or externally driven flows, would 
visit the flow domain uniformly, consistently with the geometrical
constraints imposed by its size and shape.  Any eventual nonuniformity
in the spatial particle distribution at equilibrium should
be  intrinsically associated
with the action of some external agents or forces.
This principle is applied in the next paragraph to derive
a unique representation for the
matrices $\boldsymbol{\alpha}_i({\bf x})$ and
$\boldsymbol{\beta}_i({\bf x})$.

\end{itemize}
\section{General theory}
\label{appB} 
Consider eq. (\ref{eq014}) (eq. (9) in the main text), for a fixed value of the particle coordinate ${\bf x}$. 
 Componentwise, it takes the form
\begin{eqnarray}
\frac{d v_h}{d t} & = & - \sum_{i=1}^N \sum_{k=1}^3 \frac{\alpha_{i,hk}({\bf x}) \, \lambda_i} {m} \, z_{i,k}
\,, \quad h=1,2,3 \label{seq1} \\
\frac{d z_{i,h}}{d t} & = &-\lambda_i \, z_{i,h}+ \sum_{k=1}^3  \beta_{i,hk}({\bf x}) \, v_k
+ \sqrt{2} \, \sum_{k=1}^3 c_{i,hk}({\bf x}) \, \xi_k(t) \, , \quad i=1,\dots,N\,, \;\; h=1,2,3
\nonumber
\end{eqnarray}
where $\alpha_{i,hk}({\bf x})$ $\beta_{i,hk}({\bf x})$ and $c_{i,hk}({\bf x})$ are
the entries of the matrices $\boldsymbol{\alpha}_i({\bf x})$, $\boldsymbol{\beta}_i({\bf x})$
and ${\bf c}_i({\bf x})$, respectively and $z_{i,h}$ the entries of ${\bf z}_i$.
Set 
\begin{equation}
\sigma_{hk}^i({\bf x}) = \sum_{j=1}^3 c_{i,hj}({\bf x}) \, c_{i,kj}({\bf x}) 
\,, \qquad i=1,\dots,N
\label{seq2}
\end{equation}
which, by definition, are symmetric matrices.
The associated Fokker-Planck equation for the probability density $p({\bf v},\{{\bf z}_i\}_{i=1}^N,t;{\bf x})$ reads (observe that ${\bf x}$  in the present case is not a dynamic variable but a static parameter)
\begin{eqnarray}
\frac{\partial p}{\partial t} & = & 
\sum_{h=1}^3 \frac{\partial}{\partial v_h} \left [ \sum_{i=1}^N \sum_{k=1}^3 
\frac{\alpha_{i,hk}({\bf x}) \, \lambda_i}{ m} \, z_{i,k}  \, p \right ]
+ \sum_{i=1}^N \sum_{h=1}^3 \frac{\partial}{\partial z_{i,h}}  \left ( \lambda_i \, z_{i,h} \, p \right )
\label{seq3} \\
&- & \sum_{i=1}^N \sum_{h=1}^3 \frac{\partial }{\partial z_{i,h}} \left [ \sum_{k=1}^3 \beta_{i,hk}({\bf x}) \, v_k \, p \right ]
+ \sum_{i=1}^N \sum_{h=1}^3 \sum_{k=1}^3 \frac{\partial^2}{\partial z_{i,h} \partial z_{i,k}} \left [
\sigma_{hk}^i({\bf x}) \, p \right ]
\nonumber
\end{eqnarray}
In order to enforce FD1k and FD2k moment analysis can be used, by considering the
second-order moments
\begin{eqnarray}
m_{v_h v_q}(t;{\bf x}) & = & \int v_h v_q d  {\bf v} \prod_{i=1}^N \int p( {\bf v},\{{\bf z}_i\}_{i=1}^N,t;{\bf x})
\, d {\bf z}_i \, , \quad h,q=1,2,3  \nonumber \\
m_{v_h z_{n,q}}(t;{\bf x}) & =  & \int v_h   d {\bf v} \prod_{i=1}^N  \int 
z_{n,q} \, p( {\bf v},\{{\bf z}_i\}_{i=1}^N,t;{\bf x})
\, d {\bf z}_i  \, , \quad h,q=1,2,3  \;, \quad n=1,\dots,N 
\label{seq4} \\
 m_{z_{n,h} z_{m,q}}(t;{\bf x}) & =  & \int    d {\bf v} \prod_{i=1}^N  \int z_{n,h} \, z_{m,q} \, p( {\bf v},\{{\bf z}_i\}_{i=1}^N,t;{\bf x})
\, d {\bf z}_i \; , \quad h,q=1,2,3\;, \quad n,m=1,\dots,N \nonumber
\end{eqnarray}
enforcing the conditions at equilibrium
\begin{equation}
m_{v_h \, v_k} = \frac{k_B \, T}{m} \, \delta_{hk} \;, \quad h,k=1,2,3 \; ,
\qquad
m_{v_h z_{i,k}}=0 \,, \quad h,k=1,2,3\;, \quad i=1,\dots, N
\label{seq5}
\end{equation}
The evolution  equations for the second-order moments take the form
\begin{equation}
\frac{d m_{v_p v_q}}{d t} = - \sum_{i=1}^N \sum_{k=1}^3 \frac{\alpha_{i,pk}({\bf x}) \, \lambda_i}{m} \,
m_{v_q z_{i,k}} -  \sum_{i=1}^N \sum_{k=1}^3 \frac{\alpha_{i,q k}({\bf x}) \, \lambda_i}{m} \,
m_{v_p z_{i,k}}
\label{seq6}
\end{equation}
\begin{equation}
\frac{d m_{v_p z_{n,q}}}{d t} = - \sum_{i=1}^N \sum_{k=1}^3 \frac{\alpha_{i,pk}({\bf x}) \, \lambda_i}{m} 
\, m_{z_{i,k} z_{n,q}} - \lambda_n \, m_{v_p z_{n,q}} + \sum_{k=1}^3 \beta_{n,qk}({\bf x}) \, m_{v_k v_p}
\label{seq7}
\end{equation}
\begin{equation}
\frac{d m_{z_{n,p} z_{m,q}}}{d t} = -(\lambda_n+\lambda_m) \, m_{z_{n,p} z_{m,q}}
+ \sum_{k=1}^3 \beta_{n,pk}({\bf x}) \, m_{v_k z_{m,q}} + \sum_{k=1}^3 \beta_{m,qk}({\bf x}) \, m_{v_k z_{n,p}}   + \sigma_{p,q}^n \, \delta_{nm}+ \sigma_{q,p}^n \, \delta_{nm}
\label{seq8}
\end{equation}
From eqs. (\ref{seq8}) at equilibrium (steady-state), using eqs. (\ref{seq5}),  we have
\begin{equation}
m_{z_{n,p} z_{m,q}} = \frac{\sigma_{p,q}^n({\bf x})}{\lambda_n} \, \delta_{nm}
\label{seq9}
\end{equation}
that, substituted into eq. (\ref{seq7}),  
\begin{equation}
\sum_{i=1}^N \sum_{k=1}^3 \frac{\alpha_{i,pk}({\bf x}) \, \lambda_i}{m} \, m_{z_{i,k} z_{n,q}}= \sum_{k=1}^3
\beta_{n,qk}({\bf x}) \, m_{v_k v_p}
\label{seq10}
\end{equation}
and,  using the equilibrium conditions eq. (\ref{seq5}), finally 
provide
\begin{equation}
\sum_{k=1}^3 \frac{\alpha_{n,pk}({\bf x})}{m} \, \sigma_{k,q}^n({\bf x}) = \frac{k_B \, T}{m} \,
\beta_{n,qp}({\bf x})
\label{seq11}
\end{equation}
Enforcing the symmetric nature of the matrices involved, the latter
expression can be compactly
expressed in matrix form as
\begin{equation}
\boldsymbol{\alpha}_n({\bf x}) \, \boldsymbol{\sigma}^n({\bf x}) = k_B \, T \, \boldsymbol{\beta}_n({\bf x})
\label{seq12}
\end{equation}
i.e.,
\begin{equation}
\boldsymbol{\sigma}^n({\bf x})= k_B \, T \, \boldsymbol{\alpha}_n^{-1}({\bf x}) \, \boldsymbol{\beta}_n({\bf x})
\label{seq13}
\end{equation}
The matrices $\boldsymbol{\sigma}^n({\bf x})$ are positive
definite and symmetric (since $\boldsymbol{\alpha}_n^{-1}({\bf x})$
and $\boldsymbol{\beta}({\bf x})$ possess these properties). Consequently,
for each $n$ there exists a unique symmetric and positive definite
matrix ${\bf c}_n({\bf x})$,
defined by eq.  (\ref{seq2}) such that
\begin{equation}
{\bf c}_n({\bf x})= \sqrt{k_B \, T} \, \left [ \boldsymbol{\alpha}_n^{-1}({\bf x}) \,
\boldsymbol{\beta}_n({\bf x}) \right ]^{1/2} \, , \quad n=1,\dots,N
\label{seq14}
\end{equation}
that corresponds to eq. (10) in the manuscript.

From  eq. (\ref{seq9}), it follows that the equilibrium moments $m_{z_{n,p} z_{m,q}}$ of the
auxiliary ${\bf z}_i$-variables  are proportional to
the entries $\sigma_{p,q}^n({\bf x})$ and, as a consequence,  they 
depend in general on the particle position ${\bf x}$. 

Correspondingly, the
equilibrium moments are uniform throughout the fluid domain, if and only if
the matrices $\boldsymbol{\sigma}^n$ are constant matrices. This
occurs if $\boldsymbol{\alpha}_n^{-1}({\bf x}) \, \boldsymbol{\beta}_n({\bf x})$ is
a constant matrix that, without loss of generality, can be set equal to the
identity matrix (see also a comment at the end of this paragraph). This implies
\begin{equation}
\boldsymbol{\alpha}_n({\bf x}) = \boldsymbol{\beta}_n({\bf x}) \, , \qquad n=1,\dots,N
\label{seq15}
\end{equation}
and since the product of these two matrices equals the hydromechanic matrix ${\bf A}_n({\bf x})$,
we have
\begin{equation}
\boldsymbol{\alpha}_n({\bf x})=\boldsymbol{\beta}_n({\bf x})= {\bf A}_n^{1/2}({\bf x}) \, ,
\qquad n=1,\dots,N
\label{seq16}
\end{equation}
Observe that eqs. (\ref{seq16}) are consistent with the properties
of symmetry and positive definiteness of the 
the matrices $\boldsymbol{\alpha}_n({\bf x})$ 
and $\boldsymbol{\beta}_n({\bf x})$,
inheriting these properties by
the corresponding ones  of the hydromechanic matrices ${\bf A}_n({\bf x})$.

From eqs. (\ref{seq13}), (\ref{seq16}) it follows that
\begin{equation}
\boldsymbol{\sigma}^n= k_B \, T \, {\bf I} \, , \qquad {\bf c}_n= \sqrt{k_B \, T} \, {\bf I}
\, , \qquad n=1,\dots, N
\label{seq17}
\end{equation}

Next,  consider the whole dynamic problem in which particle position ${\bf x}(t)$ is
a dynamic variable, subjected to the kinematic equations
\begin{equation}
\frac{d x_h}{d t} = v_h \, , \qquad h=1,2,3
\label{seq18a}
\end{equation}
that completes the dynamic scheme eq. (\ref{seq1}). In this
case, the Fokker-Planck equation for the density $p({\bf x},{\bf v},\{{\bf z}_i \}_{i=1}^N,t)$
associated with eqs. (\ref{seq1}),(\ref{seq18a}) is given by
\begin{eqnarray}
\frac{\partial p}{\partial t} & = & - \sum_{h=1}^3 v_h \frac{\partial p}{\partial x_h} +
\sum_{h=1}^3 \frac{\partial}{\partial v_h} \left [ \sum_{i=1}^N \sum_{k=1}^3
\frac{\alpha_{i,hk}({\bf x}) \, \lambda_i}{ m} \, z_{i,k}  \, p \right ]
+ \sum_{i=1}^N \sum_{h=1}^3 \frac{\partial}{\partial z_{i,h}}  \left ( \lambda_i \, z_{i,h} \, p \right )
\label{seq18} \\
&- & \sum_{i=1}^N \sum_{h=1}^3 \frac{\partial }{\partial z_{i,h}} \left [ \sum_{k=1}^3 \beta_{i,hk}({\bf x}) \, v_k \, p \right ]
+ \sum_{i=1}^N \sum_{h=1}^3 \sum_{k=1}^3 \frac{\partial^2}{\partial z_{i,h} \partial z_{i,k}} \left [
\sigma_{hk}^i({\bf x}) \, p \right ]
\nonumber
\end{eqnarray}
Consider the motion of the  Brownian particle in a closed and bounded domain $\Omega$ of 
volume $\mbox{meas}(\Omega)$. In this case eq. (\ref{seq18}) is equipped with reflective
boundary conditions for the velocity ${\bf v}$ at the boundary $\partial \Omega$ of $\Omega$.
It is rather straightforward to observe, that the equilibrium solution of eq. (\ref{seq18})
is uniform (position independent) if eqs. (\ref{seq16}) and (\ref{seq17})
are satisfied. In this case the equilibrium solution is given by the
generalized Maxwellian
\begin{equation}
p_{\rm eq}({\bf x},{\bf v},\{{\bf z}_i\}_{i=1}^N)=
\frac{1}{\mbox{meas}(\Omega)} \left (\frac{m}{2 \, \pi \, k_B \, T} \right )^{3/2}
e^{-m \, |{\bf v}|^2/2 \, k_B \, T}
\prod_{i=1}^N \left ( \frac{\lambda_i}{2 \, \pi \, k_B \, T} \right )^{3/2}
e^{-\lambda_i \, |{\bf z}_i|^2/2 \, k_B \, T}
\label{seq19}
\end{equation}
This follows immediately by observing that for the
Gaussian equilibrium solution eq. (\ref{seq19}) we
have
\begin{eqnarray}
\frac{\partial p_{\rm q}}{\partial x_h}=0 \, , \quad \frac{\partial p_{\rm eq}}{\partial
v_h} = - \frac{m \, v_h}{k_B \, T} \, p_{\rm eq} \, , \quad
\frac{\partial p_{\rm  eq}}{\partial z_{i,h}} = - \frac{\lambda_i \, z_{i,h}}{k_B \, T} \, p_{\rm eq} \,, \quad 
\frac{\partial^2 p_{\rm eq}}{\partial z_{i,h}\partial z_{i,k}}= \left [-\frac{\lambda_i}{k_B \, T} \, \delta_{hk} + \frac{\lambda_i^2 \, z_{i,h} \, z_{i,k}}{k_B^2 \, T^2} \right ] \, p_{\rm eq}
\label{seq20}
\end{eqnarray}
Substituting these expressions into eq. (\ref{seq18}) and enforcing 
eqs. (\ref{seq15}), (\ref{seq17}) the thesis follows.
If the matrices $\sigma^n({\bf x})$ depend on the position, it is
expected that this property, i.e. the uniformity of the marginal
spatial density, may be violated. This is confirmed by
numerical simulations reported in the manuscript, and by the
analysis of simple systems (not reported for the sake of brevity). 

In the present formulation of the thermodynamic constraints
associated with the property of equilibrium states, the
conditions $\boldsymbol{\sigma}^n({\bf x})=\mbox{const.}$, $n=1,\dots,N$, represents, at least analytically, a sufficient condition.
However, we can define the thermodynamic constraint of
spatial uniformity at equilibrium in  a stronger way: in a close
and bounded system $\Omega$, and in the absence of external perturbation
the density function associated with the dynamics
of a Brownian particle should be position independent, i.e.,
\begin{equation}
p_{\rm eq}({\bf x},{\bf v},\{ {\bf z}_i \}_{i=1}^N)= f({\bf v},\{ {\bf z}_i \}_{i=1}^N)
\label{eqxx}
\end{equation}
This stronger condition is fulfilled in a Nextonian fluid in confined geometries (eq. (5) in the manuscipt, with $U=0$) in the case the hydrodynamics
is defined by the instantaneous Stokes equations.

Adopting this stronger formulation of the spatial uniformity at equilibrium,
the density $f({\bf v},\{ {\bf z}_i \}_{i=1}^N)$ satisfies the Fokker-Planck equation
at steady state  eq. (\ref{seq3}), and consequently  following 
the analysis developed above, in order to ensure  position-independent
second-order moments,   $\boldsymbol{\alpha}_n^{-1}({\bf x}) \, \boldsymbol{\beta}_n({\bf x})$ should be necessarily constant matrices, and
eqs. (\ref{seq15})-(\ref{seq16}) follow as necessary and sufficient conditions.

A final remarks concerns the unicity of the representation eq. (\ref{seq16}).
The basic property to be enforced to ensure thermodynamic consistency is
that the products $\boldsymbol{\alpha}^{-1}_n({\bf x}) \, \boldsymbol{\beta}_n({\bf x})$ return constant matrices for any ${\bf x} \in \Omega$. 
Therefore, in principle, it may exists $N$ symmetric and positive definite
constant matrices ${\bf D}_n$ such that
\begin{equation}
\boldsymbol{\beta}_n({\bf x}) = {\bf D}_n \, \boldsymbol{\alpha}_n({\bf x})
\, , \qquad n=1,\dots,N
\label{seq21}
\end{equation}
Substituting the latter relations into the basic conditions
eqs. (\ref{eq017}) we have
\begin{equation}
\boldsymbol{\alpha}_n({\bf x}) \, {\bf D}_n \, \boldsymbol{\alpha}_n({\bf x})
= {\bf A}_n({\bf x}) \, , \qquad n=1,\dots,N
\label{seq22}
\end{equation}
for any ${\bf x} \in \Omega$. Eqs. (\ref{seq22}) admit a simple
and general solution, if and only if $\boldsymbol{\alpha}_n({\bf x})$
and ${\bf D}_n$ commute for any ${\bf x} \in \Omega$, and this
is surely ensured provided that ${\bf D}_n$ are  isotropic matrices,
i.e.
\begin{equation}
{\bf D}_n = d_n \, {\bf I} \, , \qquad n=1,\dots,N
\label{seq23}
\end{equation}
where $d_n>0$ are scalar constants. In this case
\begin{equation}
\boldsymbol{\alpha}_n({\bf x}) = d_n^{-1/2} \, {\bf A}_n^{1/2}({\bf x})
\, , \qquad 
\boldsymbol{\beta}_n({\bf x}) = d_n^{1/2} \, {\bf A}_n^{1/2}({\bf x})
\label{seq24}
\end{equation}
and this  representation implies
\begin{equation}
\boldsymbol{\sigma}_n= k_B \, T \, d_n \, {\bf I} \, , \qquad {\bf c}_n= 
\sqrt{k_B \, T \, d_n} \,{\bf I}
\label{seq25}
\end{equation}
But all these representation obtained for different values of the
positive constants $d_n$ represent equivalent stochastic Langevin equations,
and for this reason, it is convenient to assume, without loss of generality,
$d_n=1$, $n=1,\dots,N$.

\section{Dynamics in a confined Maxwell fluid}
\label{appC}

Consider the dynamics of a Brownian spherical particle of mass $m$ and radius $R_p$
in a Maxwell fluid at constant temperature $T$, confined between two parallel plates,
separated by a distance $W+2 \, R_p$. The Maxwell
fluid is
characterized by a single relaxation rate $\lambda$.
Let $x$ be the particle distance  from one of the plates,
$x \in [0,W]$. Focusing exclusively on the $x$-dynamics,
 the equations of motions attain the form
\begin{eqnarray}
\frac{d x}{d t} & = & v \nonumber \\
m \, \frac{d v}{d t}& = & - \eta_0 \, \widetilde{\eta}^p(x) \, \lambda \, z  
\label{seq2_1} \\
\frac{d z}{d t} & = & -\lambda \, z + \widetilde{\eta}^{1-p}(x) \, v + \sqrt{\frac{2 \, \widetilde{\eta}^{1-2p}(x)}{\eta_0}} \, \xi(t) \nonumber
\end{eqnarray}
where the exponent $p \in [0,1]$, $\eta_0 \, \widetilde{\eta}(x)$ represents the  Stokesian friction
in the confined system that, adopting  the mean-field approximation for the friction factor under no-slip
boundary conditions,  takes the form
\begin{equation}
\eta_0 = 6 \, \pi \, \mu \, R_p \, , \qquad \widetilde{\eta}(x)= \frac{(R_p+x) \, (R_p+W-x)}{x  \, (W-x)}
\label{sec2_2}
\end{equation}
Observe that it is possible to consider exclusively the dynamics of the
particle coordinate orthogonal to the plates because the
hydrodynamic factors depend exclusively on the particle distances
 from each of the parallel plates.

Let $L_c$, $T_c$, $V_c$ and $Z_c$ be the characteristic values for
$x$, $t$, $v$ and $z$, respectively, and introduce the nondimensional
variables
\begin{equation}
\widetilde{x}=\frac{x}{L_c} \, , \quad \widetilde{t}=\frac{t}{T_c} \, , \quad \widetilde{v}=\frac{v}{V_c} \, ,
\quad \widetilde{z}=\frac{z}{Z_c}
\label{sec2_3}
\end{equation}
Set $L_c=W$ so that $\widetilde{x} \in [0,1]$, and let 
\begin{equation}
V_c= \sqrt{\frac{k_B \, T}{m}} \, , \qquad Z_c= \sqrt{\frac{k_B \, T}{\eta_0 \, \lambda}}
\label{sec2_4}
\end{equation}
so that at equilibrim we should have $\langle \widetilde{v}^2 \rangle_{\rm eq}=\langle \widetilde{z}^2 \rangle_{\rm eq}
=1$, and define $T_c=L_c/V_c$, i.e.,
\begin{equation}
T_c= W \, \sqrt{\frac{m}{k_B \, T}}
\label{sec2_5}
\end{equation}
Let $a(\widetilde{x})= \widetilde{\eta}(W \, \widetilde{x})$,
\begin{equation}
a(\widetilde{x})= \frac{(c+ \widetilde{x}) \, (c+1-\widetilde{x})}{\widetilde{x} \, (1-\widetilde{x})}
\label{eq2_6}
\end{equation}
where $c= R_p/W$. With this notation, the non-dimensional equations of motion become
\begin{eqnarray}
\frac{d \widetilde{x}}{d \widetilde{t}} & = & \widetilde{v} \nonumber \\
\frac{d \widetilde{v}}{d \widetilde{t}} & = & - \zeta \, a^p(\widetilde{x}) \, \widetilde{z} 
\label{sec2_6} \\
\frac{d \widetilde{z}}{d \widetilde{t}} & = & - \gamma \, \widetilde{z} + \zeta \, a^{1-p}(\widetilde{x}) \, \widetilde{v}
+ \sqrt{2 \, a^{1-2 p}(\widetilde{x})} \, \varepsilon \, \xi(\widetilde{t})
\nonumber
\end{eqnarray}
where
\begin{equation}
\zeta= \frac{\eta_0 \, \lambda \, Z_c \, T_c}{m \, V_c} \,, \qquad \gamma=\lambda \, T_c \, ,
\qquad \varepsilon^2 = \frac{2 \, k_B \, T \, T_c}{\eta_0 \, Z_c^2}
\label{sec2_7}
\end{equation}
and, expliciting these expressions we have $\varepsilon^2 = \zeta$.
For notational simplicity, set $x$, $t$, $v$, and $z$ for the nondimensional variables
$\widetilde{x}$, $\widetilde{t}$, $\widetilde{v}$ and $\widetilde{z}$, respectively,
and $\alpha(x)=a^p(x)$, $\beta(x)=a^{1-p}(x)$, so that the equations
of motion attain the form
\begin{eqnarray}
\frac{d x}{d t} & = v \nonumber \\
\frac{d v}{d t} & = & - \zeta \, \alpha(x) \, z
\label{sec2_8} \\
\frac{d z}{d t} & = & - \gamma \, z + \zeta \, \beta(x) \, v + \sqrt{\frac{2 \, \beta(x)}{\alpha(x)}} \, \varepsilon \, 
\xi(t)
\nonumber
\end{eqnarray} 
Introducing the nondimensional variables
\begin{equation}
\widetilde{\lambda}= \frac{\lambda \, m}{\eta_0} \, , \qquad
\ell_{\rm ch}= W \frac{\eta_0}{m} \, \sqrt{\frac{m}{k_B \, T}}
\label{sec2_9}
\end{equation}
the parameters entering eq. (\ref{sec2_8}) can be expressed as
\begin{equation}
\gamma= \widetilde{\lambda} \, \ell_{\rm ch} \, , \qquad \zeta= \sqrt{\widetilde{\lambda}} \, \ell_{ch}\, , \qquad
\varepsilon= \sqrt{\widetilde{\lambda} \, \ell_{ch}}
\label{sec2_10}
\end{equation}

\section{Active particles in confined Maxwell fluids}
\label{appD}
Next, consider the case of an active particle in a confined Maxwell fluid. ``Activity'' is meant
in the broader sense of the occurrence of an additional fluctuational contribution in particle dynamics
that adds up to
the thermal fluctuations. Using the nondimensional formulation introduced in
the previous paragraph, the equations of motion of a ``prototypical'' active particle in
a confined geometry thus  becomes
\begin{eqnarray}
\frac{d x}{d t} & = & v \nonumber \\
\frac{d v}{d t} & =  & - \zeta \, a(x) \, z + \sqrt{2} \, \varepsilon_a \, \xi_1(t)
\label{seq3_1}
\\
\frac{d z}{ dt} & = & - \gamma \, z + \zeta \, a(x) \, v + \sqrt{2} \, \varepsilon \, \xi(t)
\nonumber
\end{eqnarray}
where $\xi_1(t)$ is the distributional derivative of a Wiener process, independent of the
Wiener process associated with $\xi(t)$, and $\varepsilon_a$  represents the intensity of the
active fluctuations. We have set $a(x)=\sqrt{A(x)}$, corresponding to the thermodynamically
consistent description of thermal fluctuations in a confined system.
In order to achieve a qualitative understanding of the dynamics, it is sufficient to consider the
motion at constant $x$, i.e. assuming the particle position as a parameter. We will show that the steady-state
 statistical 
properties are position dependent, and this is the basic  indicator of the occurrence
of   qualitative  phenomenological  differences with respect to the thermal equilibrium case of a passive particle.

The Fokker-Planck equation associated with eq. (\ref{seq3_1}) for the density $p(v,z,t;x)$  reads in
this case
\begin{eqnarray}
\frac{\partial p}{\partial t} & = & \zeta \, a(x) \, z \, \frac{\partial p}{\partial v} + \varepsilon_a^2 \, \frac{\partial^2 p}{\partial v^2} + \gamma \, \frac{\partial \left (z \, p \right )}{\partial z} 
\nonumber \\
&-& \zeta \, a(x) \, v \, \frac{\partial p}{\partial z} + \varepsilon^2 \, \frac{\partial^2 p}{\partial v^2}
\label{seq3_2}
\end{eqnarray}
The moment equations for $m_{vv}$, $m_{vz}$, $m_{zz}$ are
\begin{eqnarray}
\frac{d m_{vv}}{d t} & = & - 2 \, \zeta \, a(x) \, m_{vz} +2 \, \varepsilon_a^2 \nonumber \\   
\frac{d m_{vz}}{d t} & = & - \zeta \, a(x) \, m_{zz} - \gamma \, m_{vz}+ \zeta \, a(x) \, m_{vv}
\label{seq3_3} \\
\frac{d m_{zz}}{ d t}  & = & -2 \, \gamma \, m_{zz}+2 \, \zeta \, a(x) \, m_{vz} + 2 \, \varepsilon^2
\nonumber \\
\end{eqnarray}
At steady state, we have from the first equation,
\begin{equation}
m_{vz}= \frac{\varepsilon_a^2}{\zeta \, a(x)}
\label{seq3_4}
\end{equation}
Consequently, not only $v$ and $z$ are no longer independent (if $\varepsilon_a \neq 0$), but the
value of $m_{vz}$ depends parametrically on the particle position.
Substituting this result into the first equation at steady state we
have
\begin{equation}
m_{zz}= \frac{\zeta \, a(x) \, m_{vz}+\varepsilon^2}{\gamma}= \frac{\varepsilon_a^2 + \varepsilon^2}{\gamma}
\label{seq3_5}
\end{equation}
and this provides for $m_{vv}$ (second equation),
\begin{eqnarray}
m_{vv} & = & \frac{1}{\zeta \, a(x)} \left [ \zeta \, a(x) \, \frac{\varepsilon_a^2 + \varepsilon^2}{\gamma}
+ \frac{\gamma \, \varepsilon_a^2}{\zeta \, a(x)} \right ] \nonumber \\
& = & 1+ \varepsilon_a^2 \, \left [ \frac{1}{\gamma} + \frac{\gamma}{\zeta^2 \, a^2(x)} \right ]
\label{seq3_5a}
\end{eqnarray}
There are several implications of the latter equation: i) the second order moment for the particle
velocity depends on $x$ parametrically, and ii) its actual  steady-state value is definitely larger than
the correpsonding value, equal to $1$  characterizing passive particle statistics in  the confined system.

\end{document}